\documentclass{article}

\usepackage[preprint]{neurips_2019}

\usepackage[utf8]{inputenc} 
\usepackage[T1]{fontenc}    
\usepackage{hyperref}       
\usepackage{url}            
\usepackage{booktabs}       
\usepackage{amsfonts}       
\usepackage{nicefrac}       
\usepackage{microtype}      
\usepackage{verbatim}
\usepackage{epsfig}
\usepackage{epsf}
\usepackage{epstopdf}
\usepackage{graphicx}
\usepackage{subfigure}
\usepackage{amsmath}

\title{retina-VAE: Variationally Decoding the Spectrum of Macular Disease}

\author{%
  Stephen G. Odaibo\thanks{Correspondence:  RETINA-AI Health, Inc. P.O.Box 20169, Houston TX} \\
  (1) Department of Machine Learning Research\\RETINA-AI Health, Inc.\\
  (2) Department of Head \& Neck Surgery\\ Ophthalmology Section\\ MD Anderson Cancer Center\\
  \texttt{stephen.odaibo@retina-ai.com} \\
}

\begin{document}

\maketitle

\begin{abstract}

In this paper, we seek a clinically-relevant latent code for representing the spectrum of macular disease. Towards this end, we construct retina-VAE, a variational autoencoder-based model that accepts a patient profile vector (pVec) as input. The pVec components include clinical exam findings and demographic information. We evaluate the model on a subspectrum of the retinal maculopathies, in particular, exudative age-related macular degeneration, central serous chorioretinopathy, and polypoidal choroidal vasculopathy. For these three maculopathies, a database of 3000 6-dimensional pVecs (1000 each) was synthetically generated based on known disease statistics in the literature. The database was then used to train the VAE and generate latent vector representations. We found training performance to be best for a 3-dimensional latent vector architecture compared to 2 or 4 dimensional latents. Additionally, for the 3D latent architecture, we discovered that the resulting latent vectors were strongly clustered spontaneously into one of 14 clusters. Kmeans was then used only to identify members of each cluster and to inspect cluster properties. These clusters suggest underlying disease subtypes which may potentially respond better or worse to particular pharmaceutical treatments such as anti-vascular endothelial growth factor variants. The retina-VAE framework will potentially yield new fundamental insights into the mechanisms and manifestations of disease. And will potentially facilitate the development of personalized pharmaceuticals and gene therapies.
\end{abstract}

\section{Introduction}

In current clinical practice, physicians see patients and based on a combination of objective and subjective information gathered, make a diagnoses of the condition. The diagnosis is however typically based on a rigid and mutually exclusive classification of the disease. This system is brittle as it does not take into account that most diseases manifest between and outside of these rigid classes, and it does not provide opportunity for personalized diagnosis and treatment. Increasingly as clinicians become more aware of the inadequacies of the current system, there is a concept of ``Spectrum of disease.'' It recognizes that the genetic, metabolic, environmental, and demographic profile of a patient points to predilection for specific manifestations which may be better or less responsive to any particular treatment. In the case of the retinal maculopathies for instance, a prototypical one is age-related macular degeneration (ARMD), which in a certain subtype progresses into an exudative state which requires intravireal injections with anti vascular endothelial growth factor (anti-VEGF) injections. In the absence of this treatment, patients will often go blind. It is known to practicing retina specialists that there are multiple manifestations of this condition, simply based on the broad range of presentations and responses to treatment. However, this notion is yet to be rigorously quantified and described. For instance, certain patients with what would appear to be central serous chorioretinopathy, do progress to develop choroidal neovascular membranes and respond positively to intravitreal injections of anti-VEGF. Similarly, polypoidal choroidal vasculopathy is another related disease with clear distinctions from ARMD, that also requires intravitreal injections for treatment. In this paper, we sought out to develop a framework for the study and characteization of this notion of a ``spectrum of macular disease,'' a notion that if properly described, can potentially yield significant advances in the development of personalized pharmaceuticals and gene therapies.

From a methodological standpoint, the notion of seeking latent codes that represent a data distribution is certainly not new. Various techniques and efforts such as principal component analysis\cite{jo2011} and kmeans clustering\cite{ma1967,waca2001,cong2011,cong2012} have been ongoing for a while in the statistics community. Within machine learning, there has been some early work in adversarial generative modeling\cite{sc1992}, and more recently, a great surge of interest since the entry of generative adversarial networks \cite{gopo2014,rame2015,mios2014}. Another class, autoencoding algorithms generate a latent representation and then reconstruct the input data from the latent using the data itself as the label\cite{lera2011,hisa2006,legr2009,vila2010,ngkh2011}. These did not yield a continuous latent space, and as such were not truly generative in the sense of having capacity to generate meaningful (or interpretable) data for any point in the latent space. This was addressed by Kingma and Welling via variational autoencoders\cite{kiwe2013}, which is what we have adopted in retina-VAE. We see variational inference as particularly suitable for the problem of discovering latent features which hold diagnostic and therapeautic relevance.

There has been some related work using VAEs to determine relevant latent representations for disease stratification. For instance, Way et al used VAEs to extract a latent space from cancer transcriptomes\cite{wagr2017}. Rampasek et al used VAEs to deduce latent space of drug response in cancer cell lines. And Cohen et al used an autoencoder to obtain a representation of in situ hybridization images.\cite{cone2017}. No studies to date have yet been done on utilizing variational autoencoders to determine clinically relevant latent spaces for retinal disease diagnosis.

\section{Background}

\begin{figure}
    \centering
    \subfigure[]{\includegraphics[width=0.24\textwidth,height=0.15\textheight]{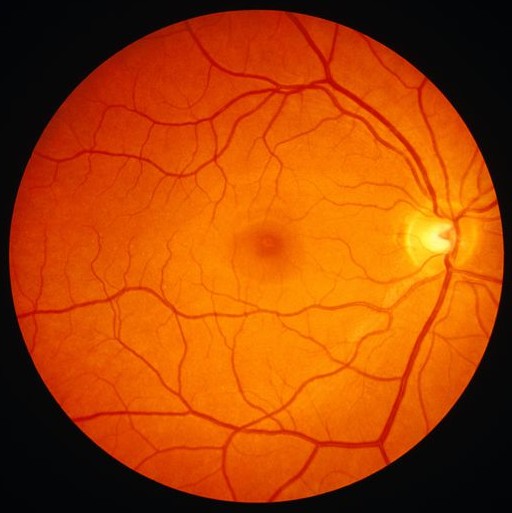}} 
    \subfigure[]{\includegraphics[width=0.24\textwidth,height=0.15\textheight]{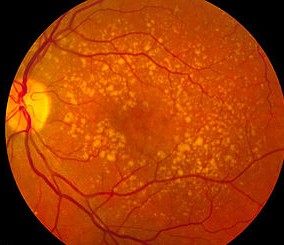}}
    \caption{(a) Normal retina (b) Retina with age-related macular degeneration showing large diffuse soft confluent drusen (yellow dots)}
    \label{fig:retina}
\end{figure}

The retina is the sensory tissue in the back of the eye. It contains the photoreceptors which absorb light and transmit a signal onwards to the primary visual cortex (area V1) via the optic nerve. Figure~\ref{fig:retina} (a) shows a normal retina. Various diseases can affect the retina resulting in vision loss and blindness. One such disease is age-related macular degeneration (ARMD)\cite{vi1989} which is the leading cause of blindness in people over the age of 50 in the United States. Figure~\ref{fig:retina} (b) shows the retina of a patient with ARMD. The condition often starts out in a ``non-exudative'' state and progresses with age into an ``exudative'' state, because of renegade blood vessels called choroidal neovascular membranes which pathologically arise from the choroid, the vascular layer underneath the retina, and break through bruch's membrane and into the subretinal space where they leak and cause vision loss. When such an exudative transformation occurs, to decrease vision loss and prevent blindness intravitreal injections of anti-VEGF are required typically anywhere from monthly to once every 3 months.

Central serous chorioretinopathy (CSCR)\cite{liqu2013,caco1992} is another disease of the macular and associated central choroid. It typically occurs in men in their 30s to 40s\cite{caco1992,tsch2013} and has been associated with stress or steroid usage. Typically, this condition resolves on its own within a few weeks and does not require particular intervention, other than a caution to abstain from any steroid use such as in nasal sprays. In its more common form, CSCR is not thought of as being related to ARMD. However, in some instances, CSCR can become chronic and can develop choroidal neovascular membranes requiring treatment with anti-VEGFs. This is typically termed ``atypical CSCR'' and is understood to be somewhere in the spectrum of ARMD. It is not currently understood why some patients with CSCR follow a different prognostic course and end up requiring intravitreal injections potentially indefinitely, while most others have complete resolution of symptoms within weeks without ever having a recurrence, and yet others have complete resolution within weeks, but then have recurrences periodically. Variational substratification of the disease and discovery of relevant latent represenattion codes will likely help answer these open questions.

Polypoidal choroidal vasculopathy (PCV) is another related condition, one in which polyps pathologically arise in the choroid. The polyps leak and can progressively disrupt bruch's membrane, causing choroidal neovascular membranes, subretinal hemorrhages (bleeds), and vision loss. The condition is often confused with exudative ARMD. Similarly to exudative ARMD, it responds positively to intravireal injections of anti-VEGF. This condition is most common in persons of Asian or African ancestry, but is increasingly being recognized in perons of European ancestry as well. PCV was not historically associated with drusen, however it is increasingly recognized that drusen-like deposits do occur in PCV patients. Such pattern is a recurrent theme undermining the traditional notion of singular `hallmark feature switches' which can absolutely discriminate between diseases. As these once presumed `absolute discriminants' are increasingly found with some, albeit lower, probability in the alternative diagnoses. This further underscores the need for probabilistic models which can distill clinically-relevant latent representation codes, and variationally infer classes of disease for personalized treatment.

In variational inference, the goal often is to determine a posterior distribution $p(z|x)$ of a latent variable $z$ given some data evidence $x$. However, determining this posterior distribution is typically computationally intractible, because according to Bayes,

\begin{equation}
 p(z|x) = \int_z \frac{p(x|z)p(z)}{p(x)}dz,
\end{equation}

which is intractible because it involves computing the integral over the entire latent space $z$, and also typically because it requires knowledge or computation of the entire evidence distribution $p(x)$. To circumvent this intractibility problem one instead approximates the posterior with some other distribution $q(z|x)$ in a manner that minimizes some similarity measure between the true posterior and the approximation, $q$. Here we use the Kullback-Leibler, $D_{KL}$:

\begin{equation}
D_{KL}(q(z|x_i)||p(z|x_i)) = \int q(z|x_i)\log \frac{q(z|x_i)}{p(z|x_i)}dz,
\end{equation}

which when manipulated in the above yields the below

\begin{equation}
 \log p(x_i) = D_{KL}(q(z|x_i)||p(z|x_i)) + E_{q(z|x_i)}\left[\log p(z,x_i)-\log q(z|x_i)\right].
\end{equation}

Since $D_{KL}\geq0$, it follows that

\begin{equation}
 \log p(x_i) \geq E_{q(z|x_i)}\left[\log p(z,x_i)-\log q(z|x_i)\right].
\end{equation}

Therefore maximizing the log likelihood of the data is synonymous with maximizing the term on the right, the variational lower bound ${\cal{L}}$. In the manner of Kingma et al~\cite{kiwe2013}, we utilize the variational lower bound as the loss function of our VAE. It can be reexpressed as:

\begin{equation}
  {\cal{L}} =  E_{q(z|x_i)}\left[\log p(z,x_i)-\log q(z|x_i)\right]= \int_z q(z|x_i)\left[\log p(z,x_i)-\log q(z|x_i)\right]dz
\end{equation}

\begin{equation}
 {\cal{L}}=\int_z q(z|x_i) \log \frac{p(z,x_i)}{q(z|x_i)}dz = \int_z q(z|x_i) \log \frac{p(x_i|z)p(z)}{q(z|x_i)}dz
\end{equation}

\begin{equation}
  {\cal{L}}= \int_z q(z|x_i) \log \frac{p(z)}{q(z|x_i)}dz + \int_z q(z|x_i) \log p(x_i|z)dz 
\end{equation}

\begin{equation}
  {\cal{L}}= -D_{KL}(q(z|x_i)||p(z)) + E_{q(z|x_i)}\left[ \log p(x_i|z)\right] 
\end{equation}

Choosing a gaussian latent prior and a gaussian approximate posterior yields a closed form for the $D_{KL}$ term. The resulting loss function is then:

\begin{equation}
 {\cal{L}}= -\frac{1}{2}\sum_{j=1}^J\left( 1 + \log(\sigma_j^2 ) - \mu_j^2 - \sigma_j^2\right) + \frac{1}{L}\sum_{l=1}^LE_{q(z|x_i)}\left[ \log p(x_i|z^{(i,l)}) \right]
 \label{Eq:vae_loss}
\end{equation}

where $J$ is the dimension of the latent, $\sigma_j$ and $\mu_j$ are parameters of the approximate posterior, $q$, and $L$ is the number of samples stochastically drawn utilizing the reparametrization trick\cite{kiwe2013}.

\begin{figure}
    \centering
     \subfigure[]{\includegraphics[width=0.24\textwidth]{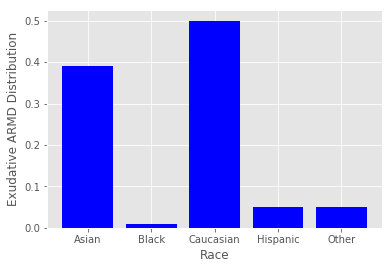}}
    \subfigure[]{\includegraphics[width=0.24\textwidth]{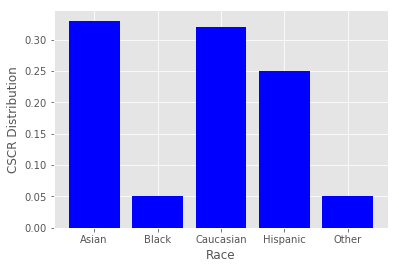}} 
     \subfigure[]{\includegraphics[width=0.24\textwidth]{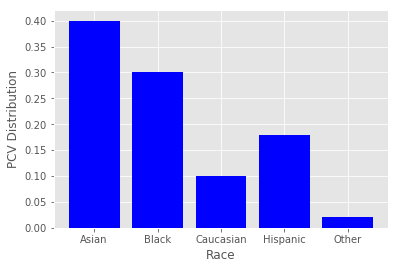}} \\
          \subfigure[]{\includegraphics[width=0.24\textwidth]{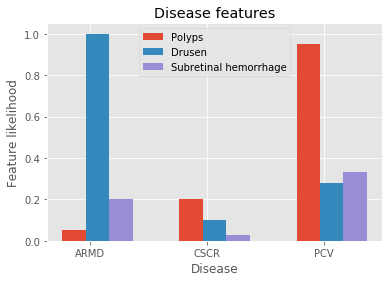}} 
                    \subfigure[]{\includegraphics[width=0.24\textwidth]{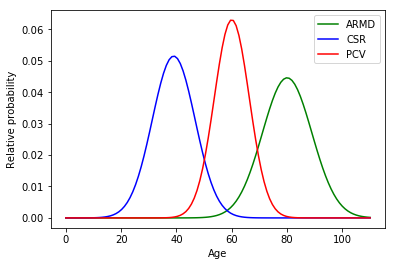}} 
    \caption{(a) Racial Distribution of Exudative Age-Related Macular Degeneration (ARMD) (b) Racial Distribution of Central Serous Chorioretinopathy (CSCR) (c) Racial Distribution of Polypoidal Choroidal Vasculopathy (PCV) (d) Distribution of polyps, drusen, and subretinal hemorrhage in ARMD, CSR, and PCV (e) Age Distributions of exudative ARMD, of CSCR, and of PCV.}
    \label{fig:pVec_data}
\end{figure}

\begin{figure}[h]
\begin{center}
\scalebox{.5}
{\includegraphics{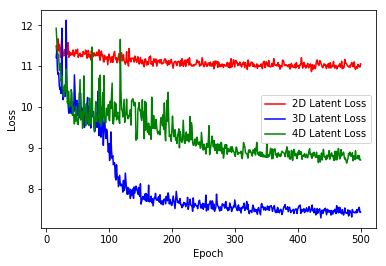}}
\end{center}
\caption[mid]{Training Spectral-VAE with 2, 3, and 4 dimensional latent}
\label{fig:training_234}
\end{figure}

\begin{figure}[h]
\begin{center}
\scalebox{.5}
{\includegraphics{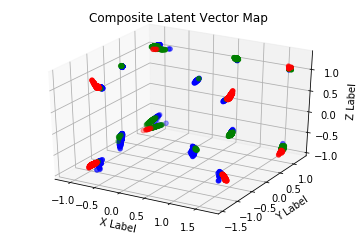}}
\end{center}
\caption[mid]{Composite Latent Vector Map}
\label{fig:compo_latent}
\end{figure}

\begin{figure}
    \centering
    \subfigure[]{\includegraphics[width=0.24\textwidth]{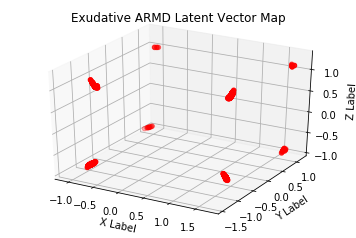}} 
    \subfigure[]{\includegraphics[width=0.24\textwidth]{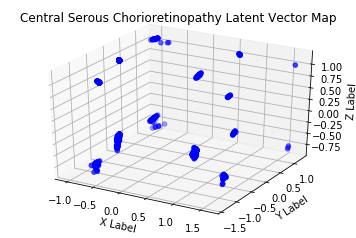}} 
    \subfigure[]{\includegraphics[width=0.24\textwidth]{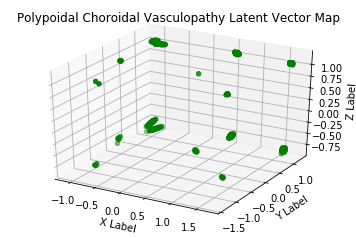}}\\
    \subfigure[]{\includegraphics[width=0.24\textwidth]{latent_z3.png}}
    \caption{(a) 3D latent vector space of exudative ARMD (b) 3D latent vector space of CSCR (c) 3D latent vector space of PCV (d) 3D latent vector space composite of exudative ARMD, CSCR, and PCV}
    \label{fig:3D_latents}
\end{figure}

\begin{figure}
    \centering
    \subfigure[]{\includegraphics[width=0.24\textwidth]{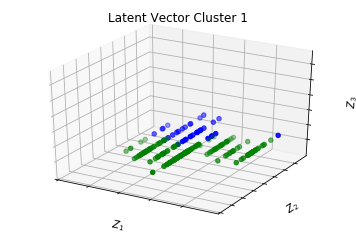}} 
    \subfigure[]{\includegraphics[width=0.24\textwidth]{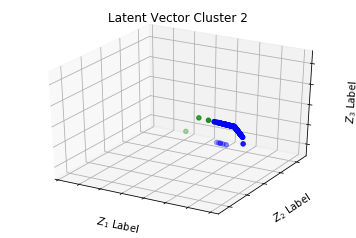}} 
    \subfigure[]{\includegraphics[width=0.24\textwidth]{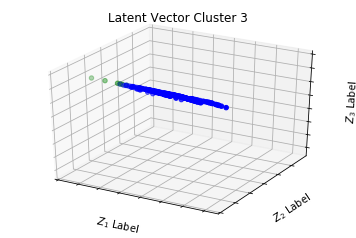}}
    \subfigure[]{\includegraphics[width=0.24\textwidth]{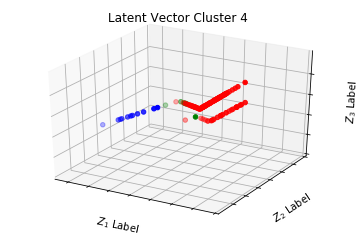}}\\
        \subfigure[]{\includegraphics[width=0.24\textwidth]{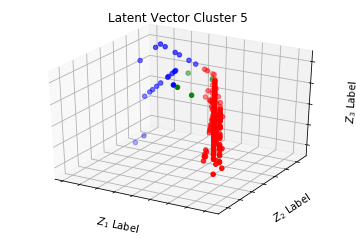}} 
    \subfigure[]{\includegraphics[width=0.24\textwidth]{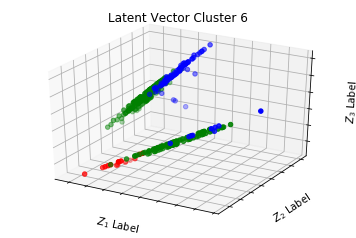}} 
    \subfigure[]{\includegraphics[width=0.24\textwidth]{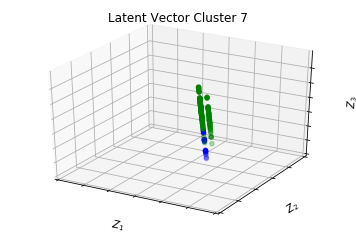}}
    \subfigure[]{\includegraphics[width=0.24\textwidth]{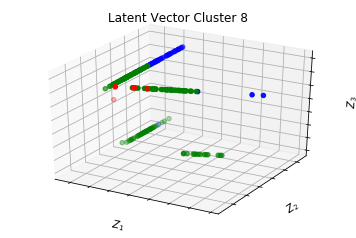}}\\
    \subfigure[]{\includegraphics[width=0.24\textwidth]{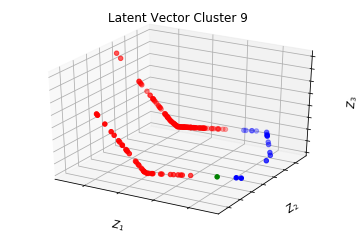}} 
    \subfigure[]{\includegraphics[width=0.24\textwidth]{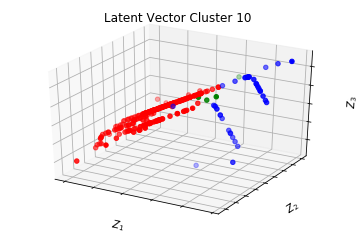}} 
    \subfigure[]{\includegraphics[width=0.24\textwidth]{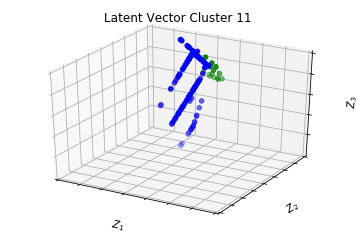}}
    \subfigure[]{\includegraphics[width=0.24\textwidth]{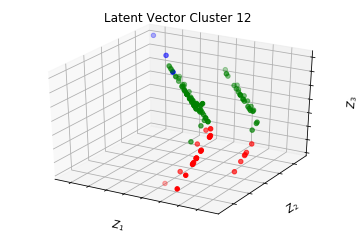}}\\
        \subfigure[]{\includegraphics[width=0.24\textwidth]{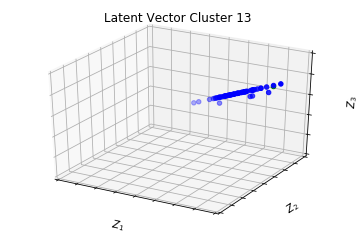}} 
    \subfigure[]{\includegraphics[width=0.24\textwidth]{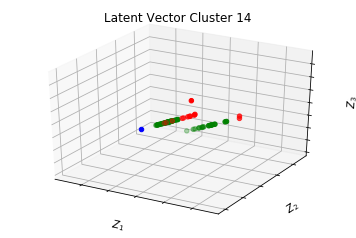}} 
    \caption{(a) Cluster 1 (b) Cluster 2 (c) Cluster 3 (d) Cluster 4 (e) Cluster 5 (f) Cluster 6 (g) Cluster 7 (h) Cluster 8 (i) Cluster 9 (j) Cluster 10 (k) Cluster 11 (l) Cluster 12 (m) Cluster 13 (n) Cluster 14}
    \label{fig:Clusters}
\end{figure}

\section{Methods}

Three thousand patient profile vectors (pVec) were generated based on literature on the epidemiology of the three maculopathies under study. Mean values and distribution information was obtained, distribution models built, and then sampling carried out on the distributions. The pVec consisted of the following 6 components: race, age, presence or absence of polyps, presence or absence of drusen or drusen-like deposits, presence or absence of subretinal hemorrhage (SRH), and sex. The random variables polyps, drusen, SRH, and sex had values of either 0 or 1. The race categories were asian, black, caucasian, hispanic, and other. The age random variable was continuous, assuming values in the positive reals. 

\subsection*{Data Model}
For the data model, it was noted that CSCR has a median age of about 36 to 39 years\cite{caco1992,tsch2013}. Hence for the CSCR age distribution model, we used a mean age of 39 years and a variance of 60. The mean age of patients affected with PCV varied by study between 55 and 68 years \cite{shota2003,yaso2012}. Hence for the PCV age distribution model we used a mean age of 60 years and a variance of 40. The average age of patients with exudative ARMD is about 80 years \cite{johu2003}. Hence for exudative ARMD we used a mean age of 80 years and a variance of 80 in the data model. Figure~\ref{fig:pVec_data} (e) shows age distribution of exudative ARMD, of CSCR, and of PCV used in our data model. Notably, exudative ARMD is most common amongst asians and caucasians and least common amongst blacks and hispanics~\cite{frbr1999,kaya2010,vaza2011,klkl2006}. Figure~\ref{fig:pVec_data} (a) shows the racial distribution of exudative ARMD used in our data model. PCV is notably more common in people of Asian and African descent than in other racial groups~\cite{cido2004,imen2010}, while CSCR is most common in asians, caucasians, and hispanics, and is relatively rare in blacks~\cite{liqu2013,deal2003,ahge2009,lich2016}. Figures~\ref{fig:pVec_data} (b) and (c) respectively show the racial distribution of CSCR and of PCV used in our data model. In exudative ARMD, CSCR, and PCV, we respectively used [0.39, 0.01, 0.5, 0.05,0.05], [0.33, 0.05, 0.32, 0.25, 0.05], and [0.4, 0.3, 0.10, 0.18, 0.02] as the categorical distributions amongst asians, blacks, caucasians, hispanics, and other races respectively. Drusen and drusen-like deposits have been observed in up to one third of patients with PCV~\cite{lale2000,iwts2008}. In our model we use 28\% coincidence of drusen and PCV. Drusen or drusen-like deposits are known to occur in CSCR~\cite{scma1992,chfr1974} even though this is not a common occurence. We model with 10\% drusen or drusen-like occurence conincident with CSCR. About one third of patients with PCV were found to have large subretinal hemorrhage in one study~\cite{byle2008}. No definite sex predilection has been established for PCV~\cite{yado1999}. Neither has there been any definite sex predilection established for exudative ARMD~\cite{vi1989}. Of note, for exudative ARMD a few studies have noted some potentially race-specific sex predilection in either direction\cite{mali2007,ruwo2012}, but nothing definite across the board. In our data model we used a uniform bernouli distribution for sex of PCV and exudative ARMD patients. In CSCR, there is a significant male preponderence ranging anywhere from 1:3 to 1:7 ~\cite{roba2016,kipu2008}. Our dataset has been made publicly available for download at https://github.com/VeryVeryGoodNews/retina-VAE.

\subsection*{Model Architecture}
The VAE was implemented in Keras using the loss function described in Equation~\ref{Eq:vae_loss} above. The model architecture's encoder part consisted on 6-dimensional input vector, one hidden layer of size 512, and latent vectors of varying dimensions (2, 3, and 4) were set in experiment runs. For the encoder network, a \textit{reLu} was used as the activation function of the hidden layer, and a \textit{linear} activation function was used in the output. For the decoder network, a \textit{reLu} was used for the activation function of the hidden layer, while a \textit{sigmoid} was used in the output layer.  The decoder part of the model also consisted of a single hidden layer of size 512 and a 6 dimensional output. Training was done on an NVIDIA Tesla V100 GPU with 16 GB RAM. And the number of iterations on the VAE was set to 1000 epochs. The reparametrization trick\cite{kiwe2013} was utilized for the stochastic sampling of the latent vector during training. The loss function was plotted to monitor training. The code is available for download at  https://github.com/VeryVeryGoodNews/retina-VAE.

\begin{table}
  \caption{\textbf{Cluster characteristics}}
  \label{tab:cluster_characteristics}
  \centering
  \begin{tabular}{c||c|c|c|c|c|c|c}
    \toprule
    \cmidrule(r){1-2}
    ID & Size    & Race  & Age     & Polyps  & Drusen  & SRH  & Sex  \\
    \hline
    \hline
     1 & 249 & [249,0,0,0,0] & [26,56,86] & [249,0] & [189,60] & [68,181] & [249,0]\\\hline
     2 & 198 & [0,35,163,0,0]  & [23,39,70] & [0,198] & [0,198] & [8,190] & [198,0]\\\hline
     3 & 298 & [0,8,229,35,0] & [23,79,104] & [0,298] & [298,0] & [57,241] & [0,298]\\\hline
     4 & 220 & [220,0,0,0,0] & [24,79,105] & [0,220] & [220,0] & [41,179] &[220,0] \\\hline
     5 &  339& [0,141,78,104,16] & [21,60,104]  & [339,0] & [112,227] & [101,238] & [0,339]\\\hline
 6 & 225 & [225,0,0,0,0] & [30,61,92]  & [225,0] & [71,154] & [66,159] & [0,225]\\\hline
 7& 203 & [0,129,74,0,0] & [20,57,78]  & [203,0] & [37,166] & [55,148] & [203,0]\\\hline
 8 & 186 & [186,0,0,0,0] & [30,78,107]  & [0,186] & [186,0] & [32,154] & [0,186]\\\hline
 9 & 348 & [0,11,268,43,26] & [22,79,106]  & [0,348] & [348,0] & [54,294] & [348,0]\\\hline
 10 & 154 & [0,17,60,61,16] & [14,40,75]  & [0,154] & [0,154] &[8,146]  &[0,154]\\\hline
 11 & 165 & [165,0,0,0,0] & [22,40,75]  & [0,165] & [0,165] & [7,158]& [165,0]\\\hline
 12 & 69 &  [69,0,0,0,0] & [22,39,59]  & [0,69] & [0,69] &[4,65]  & [0,69]\\\hline
 13 & 195 & [0,0,36,141,18] & [25,60,91]  & [195,0] & [77,118] & [61,134] &[195,0] \\\hline
 14 & 151 & [0,0,0,134,17] & [18,39,63]  & [0,151] & [0,151] & [11,140] &[151,0] \\
    \bottomrule
  \end{tabular}
  \caption{The `Size' col is number of patients in cluster. `Race' col shows race distribution with notation: [Asian, Black, Caucasian, Hispanic, Other]. `Age' column shows the age distribution as [min age, median age, max age]. `Polyps,' `Drusen,' and `SRH' cols show respective distributions as [number with, number without]. And `Sex' col shows sex distribution as [number male, number female].}
\end{table}

\section{Experiments \& Results}

After training the VAE, the encoder network was detached such that the latent vector, $z$ was encoder output. Inference was then run on this trained encoder to determine the latent vectors that corresponded to each pVec vector in the training set. The latent vectors were plotted in the 3D latent case and 14 spontaneously formed clusters were visually observed as shown in Figure~\ref{fig:3D_latents}.  Upon observation of these 14 clusters, kmeans was run with a cluster size set to 14, so as to identify the members of the individual clusters. The overlay of latent vectors of all 3 maculopathies under study is shown in Figure~\ref{fig:3D_latents} (d). Notably, 8 of the 14 clusters were found to contain representatives from all 3 maculopathies under study. Those 8 are clusters 4, 5, 6, 8, 9, 10, 12, and 14. On the other hand, 6 of the 14 clusters were found to contain representatives from only CSCR and PCV. Those 6 clusters are clusters 1, 2, 3, 7, 11, and 13. The sizes of the clusters varied as shown in Table~\ref{tab:cluster_characteristics}. The `ID' column shows the cluster ID, i.e cluster 1 has ID of 1. The `Size' column shows number of patients in the cluster. The `Race' column shows the race distribution in the cluster, and we have used notation [Asian, Black, Caucasian, Hispanic, Other]. The `Age' column shows the age distribution of the cluster, shown as [min age, median age, max age]. The `Polyps' column shows the polyps distribution of patients in the cluster, shown as [number with, number without]. The `Drusen' column shows the `drusen' distribution of patients in the cluster, shown as [number with, number without]. The `SRH' column shows the subretinal hemorrhage distribution in the cluster, shown as [number with, number without]. And the `Sex' column shows the sex distribution in the cluster, shown as [number male, number female]. Interestingly, the polyps and sex attributes were not mixed between clusters in the sense that for any given cluster, the patients in the cluster either all had polyps or none of them had polyps; also, they were either all male or all female. Also interesting was the number of clusters which had all Asians and no other races. Those clusters are 1, 4, 6, 8, 11, and 12. No other races were the sole occupants of any cluster. Clusters 2, 12, and 14 had the lowest median ages of 39. Clusters 10 and 11 also had young median ages of 40. Clusters 3, 4, and 9 had the oldest median ages of 79, and cluster 8 was close with a median age of 78. Subretinal hemorrhage was the only categorical random variable that was completely mixed amongst the clusters, i.e. all clusters had some but not all patients with SRH.

\section{Discussion}

The determination of clinically-relevant latent codes holds the key to personalized medicine, and represents a critical step in that direction. It is interesting to ask how retina-VAE mechanistically went about grouping the pVecs into these 14 classes. In statistical machine learning, dimensionality reduction is often presented as a process in which irrelevant information is discarded. However, our result here makes it apparent that dimensionality reduction can be an intelligent and more efficient rearrangement of information to fit into smaller dimensions. In this sense, one can think of the 14 clusters as representing an additional coordinate in the address space of the disease representation. For instance, such that when the decoder network “reads” a latent vector as arising from cluster 4 and knows to reconstruct that pVec as being of the Asian race and male sex. However, there are 14 spontaneous clusters yet only 4 or less of any given attribute in the pVec. Therefore, clearly as shown in Table~\ref{tab:cluster_characteristics}, most clusters have a mixture of multiple different categories of any given random variable. And these clusters are likely to have biological relevance and hold insights into the mechanisms of macular disease. 

The concept of dimensionality reduction such as going from a 6 dimensional pVec to a 3D latent space is clinically appealing as it agrees with our intuition and clinical practice. It agrees with the notion that common pathophysiology and therefore common pharmacology connects and determines diseases in a spectrum. This concept of a spectrum of diseases is what we have set out here to decode for the maculopathies. Notably, all three of the maculopathies under study have forms or states that require treatment with anti-VEGF drugs. Exudative ARMD and exudative PCV are treated with anti-VEGF in common and standard clinical practice, while atypical CSCR is a less common state often considered (or confused with) a variant of exudative ARMD. All three conditions pathophysiologically involve or can progress towards a disruption of Bruch’s membrane and the subsequent formation of a choroidal neovascular membrane. Once in this state, all three of these maculopathies respond positively to intravitreal injection of anti-VEGF. This all suggests common physiological pathways which converge under certain circumstances. 

As shown in Figure~\ref{fig:Clusters}, within each of the 14 clusters, the individual maculopathies are visually and computationally separable. Thus suggesting a multidimensional spectrum that could yield fundamental insights about which patients will respond better or worse to what drugs. And consequently, about how to develop the best drugs.

\section{Conclusion}
We have introduced, retina-VAE, a novel application of variational autoencoders towards determining a clinically relevant latent space for exudative macular degeneration and related conditions which require treatment with anti-VEGF agents. The latent vectors spontaneously formed into 14 clinically-relevant clusters which are candidates for further disease characterization, personalized intervention, and optimal pharmaceutical development.

\section{Future Work}

The authors plan to further characterize and study the 14 clusters which formed spontaneously in retina-VAE. A clinical study will be conducted in which patients at our health center will be prospectively classified into one of the 14 clusters by running inference using the trained retina-VAE model. These patients will then be followed to assess their phenotypic characteristics such as disease course, response to anti-VEGF treatment, injection interval, genetic profile, co-morbidities, family history, and environmental exposures. This information could then potentially be used to provide treatment recommendations to optimize patient outcomes and to facilitate new drug development.

\subsubsection*{Acknowledgments}

The author thanks Google Cloud for providing the compute resources used in the experiments.

\medskip

\small


\end{document}